\documentclass[9pt,sigconf]{acmart}
\renewcommand\footnotetextcopyrightpermission[1]{}
\usepackage{threeparttable}
\usepackage{textcomp}
\usepackage[table]{xcolor}

\usepackage{comment}
\usepackage{booktabs}
\usepackage{multirow}
\usepackage{array}
\usepackage[labelformat=simple]{subcaption}

\usepackage{amsmath}   

\usepackage[mathcal]{eucal}
\usepackage[ruled,vlined,noend]{algorithm2e}
\usepackage[]{algpseudocode}                               
\usepackage{enumitem} 
\usepackage{cleveref}
\Crefformat{figure}{Figure~#2#1#3}                           
\Crefname{subfigure}{Figure}{Figs.}
\Crefname{figure}{Figure}{Figs.}
\Crefformat{table}{Table~#2#1#3}
\usepackage{tabularx}

\usepackage{multirow}
\usepackage{xcntperchap}
\usepackage{tikz}

\usepackage{fancyhdr} 
\AtBeginDocument{%
  }

\setcopyright{acmlicensed}
\copyrightyear{2018}
\acmYear{2018}
\acmDOI{XXXXXXX.XXXXXXX}
\acmConference[Conference acronym 'XX]{Make sure to enter the correct
  conference title from your rights confirmation email}{June 03--05,
  2018}{Woodstock, NY}
\acmISBN{978-1-4503-XXXX-X/2018/06}

\begin{document}

\title{RTLSeek: Boosting the LLM-Based RTL Generation with Multi-Stage Diversity-Oriented Reinforcement Learning}

 %


\author{Xinyu Zhang}
\authornotemark[1] 
\affiliation{%
  \institution{State Key Lab of Processors, Institute of Computing Technology, Chinese Academy of Sciences}
  \country{China}
}
\affiliation{%
  \institution{University of Chinese Academy of Sciences}
  \country{China}
}

\author{Zhiteng Chao}
\authornote{Both authors contributed equally to this research.}
\affiliation{%
  \institution{State Key Lab of Processors, Institute of Computing Technology, Chinese Academy of Sciences}
  \country{China}
}

\author{Yonghao Wang}
\affiliation{%
  \institution{State Key Lab of Processors, Institute of Computing Technology, Chinese Academy of Sciences}
  \country{China}
}

\author{Bin Sun}
\affiliation{%
  \institution{State Key Lab of Processors, Institute of Computing Technology, Chinese Academy of Sciences}
  \country{China}
}
\affiliation{%
  \institution{University of Chinese Academy of Sciences}
  \country{China}
}

\author{Tianyun Ma}
\affiliation{%
  \institution{State Key Lab of Processors, Institute of Computing Technology, Chinese Academy of Sciences}
  \country{China}
}
\affiliation{%
  \institution{University of Chinese Academy of Sciences}
  \country{China}
}

\author{Tianmeng Yang}
\affiliation{%
  \institution{Peking University}
  \country{China}
}

\author{Jianan Mu}
\authornote{Corresponding author: mujianan@ict.ac.cn.}
\affiliation{%
  \institution{State Key Lab of Processors, Institute of Computing Technology, Chinese Academy of Sciences}
  \country{China}
}

\author{Jing Justin Ye}
\affiliation{%
  \institution{State Key Lab of Processors, Institute of Computing Technology, Chinese Academy of Sciences}
  \country{China}
}
\affiliation{%
  \institution{University of Chinese Academy of Sciences}
  \country{China}
}
\affiliation{%
  \institution{CASTEST Co., Ltd.}
  \country{China}
}

\author{Huawei Li}
\authornote{Corresponding author: lihuawei@ict.ac.cn.}
\affiliation{%
  \institution{State Key Lab of Processors, Institute of Computing Technology, Chinese Academy of Sciences}
  \country{China}
}
\affiliation{%
  \institution{University of Chinese Academy of Sciences}
  \country{China}
}
\affiliation{%
  \institution{CASTEST Co., Ltd.}
  \country{China}
}

\renewcommand{\shortauthors}{Chao and Zhang, et al.}


\begin{abstract}

Register Transfer Level (RTL) design translates high-level specifications into hardware using HDLs such as Verilog. Although LLM-based RTL generation is promising, the scarcity of functionally verifiable high-quality data limits both accuracy and diversity. Existing post-training typically produces a single HDL implementation per specification, lacking awareness of RTL variations needed for different design goals. 
We propose RTLSeek, a post-training paradigm that applies rule-based Diversity-Oriented Reinforcement Learning to improve RTL correctness and diversity. Our Diversity-Centric Multi-Objective Reward Scheduling integrates expert knowledge with EDA feedback, and a three-stage framework maximizes the utility of limited data. Experiments on the RTLLM benchmark show that RTLSeek surpasses prior methods, with ablation results confirming that encouraging broader design-space exploration improves RTL quality and achieves the principle of ``the more generated, the better results.'' 
Implementation framework, including the dataset, source code, and model weights, is shown at~\url{https://anonymous.4open.science/r/DAC2026ID71-ACB4/}.



\end{abstract}

\maketitle
\section{Introduction}

Register Transfer Level (RTL) design translates high-level functional descriptions into circuit block architectures and remains the most complex stage of digital hardware customization. Unlike downstream logic and physical design, RTL coding is still poorly automated and heavily dependent on engineer expertise.
Recent advances in Large Language Models (LLMs)~\cite{brown2020language,ouyangTrainingLanguageModels,openaiGPT4TechnicalReport2024} have enabled automatic code generation across many programming languages, motivating research on LLM-driven RTL generation. Early work shows that Supervised Fine-Tuning (SFT) can align human language with RTL design languages and functional specifications~\cite{changDataAllYou2024a,chenDawnAINativeEDA2024,liuChipNeMoDomainAdaptedLLMs2024,thakurBenchmarkingLargeLanguage2023,liuVerilogEvalEvaluatingLarge2023,luRTLLMOpenSourceBenchmark2023}.
However, progress remains limited: even on simple benchmarks, RTL generation accuracy stays below 60\%~\cite{liuDEEPRTLBRIDGINGVERILOG2025}. A primary bottleneck is the scarcity of high-quality training data—only about one thousand examples include testbenches, which are essential for verifying correctness. From an academic standpoint, a substantial expansion of open-source, functionally verifiable chip design data appears unlikely in the near term.
This raises a fundamental question for the community:
\textbf{How can we leverage circuit design expertise to improve LLM training without relying on more testbench-equipped datasets?}

\begin{figure}[tb!]
\center
\includegraphics[width=0.75\linewidth]{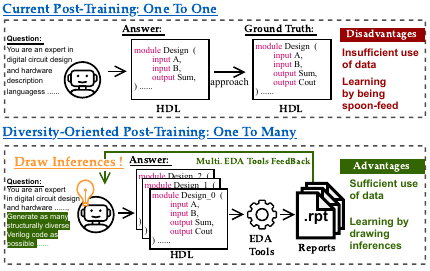}
\vspace{-0.2cm}
\caption{Motivation of RTLSeek: diversity-oriented post-training alleviates data scarcity limitations in current SFT-based RTL generation.}
\vspace{-0.4cm}
\label{fig1}
\end{figure}

Human learners often approach RTL design through \textbf{``learning by seeking''}: exploring multiple solutions for a single problem, validating them, and extracting deeper insights from limited examples. In contrast, as illustrated in \Cref{fig1}, the widely used one-to-one post-training paradigm (e.g., SFT) encourages rote memorization rather than genuine reasoning about RTL design principles.

Inspired by human learning, we propose encouraging LLMs to actively explore diverse RTL implementations—generating multiple structural variants for the same description and evaluating them—thereby maximizing the utility of scarce testbench-equipped data. Achieving this requires overcoming two challenges:
\textcircled{1} Training efficiency: enabling diverse exploration with a limited dataset;
\textcircled{2} Diversity–correctness balance: ensuring rigorous functional correctness while broadening design space exploration.
Recent progress, such as DeepSeek R1~\cite{deepseek-aiDeepSeekR1IncentivizingReasoning2025}, demonstrates that Group Relative Policy Optimization (GRPO)-based Reinforcement Learning (RL) can unlock emergent reasoning abilities in LLMs, reshaping post-training methodologies.

Motivated by human learning and these methodological advances, we introduce \textbf{RTLSeek}, a diversity-oriented RL framework for LLM-based RTL code generation. To maximize scarce training data, RTLSeek integrates SFT with a two-stage GRPO process. We design a specialized reward mechanism—automatically generated via circuit analysis tools—to balance correctness learning and diversity exploration. To further promote structurally diverse design attempts, we compute diversity rewards through Abstract Syntax Tree (AST)–based structural equivalence analysis.

Our contributions are summarized as follows:

    
    


\begin{itemize}
\item We present RTLSeek, a diversity-oriented reinforcement learning framework for LLM-based RTL generation. By integrating GRPO into training, RTLSeek systematically explores structural diversity and substantially improves RTL code generation quality.

\item To balance diversity and correctness, RTLSeek employs a \textbf{Multi-Objective Reward Scheduling} strategy that combines expert IC design constraints, EDA tool feedback, and dynamic assessments of intermediate-generation quality.

\item To address the scarcity of verifiable datasets, RTLSeek introduces a \textbf{three-stage training pipeline}: (1) SFT warm-up, (2) diversity-oriented exploration, and (3) multi-objective exploration, achieving an effective trade-off between exploration depth and functional correctness.

\item Experimental results show that our post-training paradigm improves Qwen 2.5’s RTL generation performance by over 40\%, surpassing other methods. Ablation studies confirm that removing diversity rewards or any training stage degrades performance, highlighting the effectiveness of our diversity-oriented RL approach in data-scarce RTL tasks.

\end{itemize}
\section{Observation}

Test Time Scale (TTS) technology is usually used to enhance generation ability of LLMs through the strategic design of user prompts as shown in~\Cref{fig1}. For more challenging generation tasks, by adjusting input prompts during the test phase, the same LLM is able to generate multiple structurally diverse RTL code snippets, among which a functionally correct module is more likely to emerge. If the LLM is only asked to produce a single output, it becomes much harder to generate the correct result in one attempt.

This observation also highlights a key distinction: unlike software code, RTL exhibits inherent concurrency, where even minor structural changes can significantly impact functional correctness. This raises the question of whether enhancing the model's reasoning and RTL understanding during training—so as to generate structurally diverse RTL modules—could improve the success rate of LLMs in RTL generation tasks, ultimately advancing toward the goal of \textbf{“the more generated, the better the results.”}

\begin{figure}[t!]
    \center
    \includegraphics[width=0.75\linewidth]{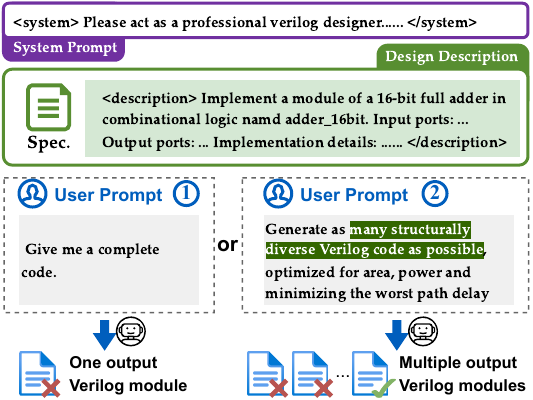}
    \vspace{-0.2cm}  
    \caption{Observation from GPT-4's RTL design: achieving Test Time Scale (TTS)~\cite{zhang2025and} by designing a multi-sample user prompt enhances the correctness of generated RTL syntax and functionality.}
    \label{fig1}
    \vspace{-0.4cm}  
\end{figure}


\section{Design of RTLSeek}



\begin{figure*}[htb!]
    \center
    \includegraphics[width=0.9\linewidth]{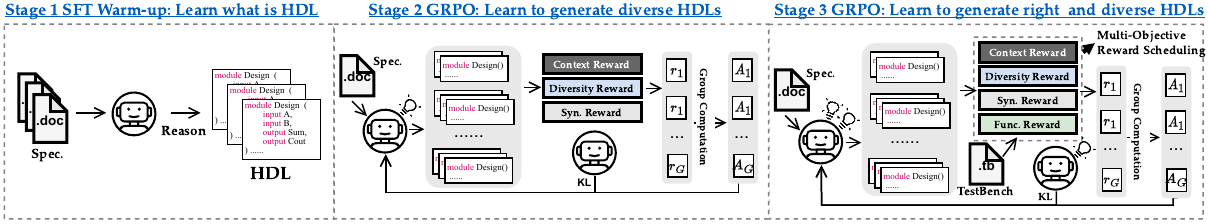}
    \vspace{-0.4cm}  
    \caption{The three-stage hybrid training paradigm in RTLSeek.}
    \vspace{-0.4cm}  
    \label{fig2}
\end{figure*}


In this section, we present the design of RTLSeek as shown in \Cref{fig2}.
First, we formalize the problem formulation for LLM-based RTL generation and outline the overall design of RTLSeek. 
Next, we detail our hybrid training paradigm that integrates SFT with GRPO-based reinforcement learning, followed by the multi-objective reward mechanism that dynamically balances correctness and diversity. Finally, we present our AST-based structural analysis approach for diversity quantification.
\subsection{Problem Definition}\label{subsec:met_pro}
Our problem can be defined as follows.
Given a natural language specification $S$ describing desired hardware behavior (see \emph{user prompt} in \Cref{fig3}), the RTL generation task aims to produce a set of design implementations $\{D_1, D_2, ..., D_n\}$ satisfying:

\begin{enumerate}[leftmargin=*,align=left]
    \item \textbf{Functional Equivalence}: 
    $\forall D_i, D_j \in \{D\}: D_i \equiv D_j$ where $\equiv$ denotes behavioral equivalence verified through testbench.
    \item \textbf{Structural Diversity}: 
    $\operatorname{Struct}(D_i) \neq \operatorname{Struct}(D_j)$ for $i \neq j$, where $\operatorname{Struct}(\cdot)$ represents the Abstract Syntax Tree (AST) representation.
    \item \textbf{Non-trivial Variation}: 
    $\operatorname{Diversity}(\operatorname{Struct}(D_i), \operatorname{Struct}(D_j)) $ $> \delta$, where $\delta$ is the minimum variation threshold excluding superficial changes (e.g., variable renaming).
\end{enumerate}

\subsection{Overall Design}\label{subsec:met_ovr}
This paper addresses a critical problem in RTL generation using LLMs: how to effectively leverage existing circuit design expertise to enhance model training, given the severe scarcity of high-quality, verifiable training data. 
The current landscape presents a fundamental limitation that only several thousand samples with accompanying testbenches (essential for validating model-generated RTL code) are publicly available. 
This data scarcity creates a pressing need for alternative training paradigms that can compensate for the lack of large-scale, high-quality datasets.

Existing solutions fall short in two key aspects. First, verification-based RTL generation methods~\cite{sami2024aivril,huang2024towards} achieve only superficial diversity improvements (e.g., simple variable renaming) at high computational cost.
Second, while SFT ensures basic correctness, it fails to produce sufficiently diverse designs; conversely, pure diversity-driven approaches often generate non-functional code.
This reveals a fundamental tension in RTL generation: how to simultaneously ensure functional correctness while exploring meaningful design variations under tight data constraints.

To address these challenges, we propose RTLSeek, a GPRO-based training paradigm that combines three key innovations:
\begin{itemize}
    \item A hybrid training paradigm integrating SFT with two-stage GRPO optimization (as shown in \Cref{fig2}); 
    \item An automated reward system using circuit analysis tools to balance correctness and diversity (as shown in \Cref{fig3}); 
    \item AST-based structural equivalence analysis to quantify and encourage design diversity (as shown in \Cref{fig4}).
    
\end{itemize}

Unlike previous methods that treat correctness and diversity as competing objectives, RTLSeek's integrated approach enables efficient exploration of the design space while maintaining functional validity, even with limited training data.

\subsection{The Hybrid Training Paradigm}\label{subsec:met_hyb}
In this section, we introduce the hybrid training paradigm integrating the SFT and the two-stage GRPO-based RL.
We first present the preliminary of the GRPO-based RL.
Then we give the details of the training paradigm.

\subsubsection{GPRO-based Reinforcement Learning}

\begin{figure*}[tb!]
    \center
    \includegraphics[width=0.75\linewidth]{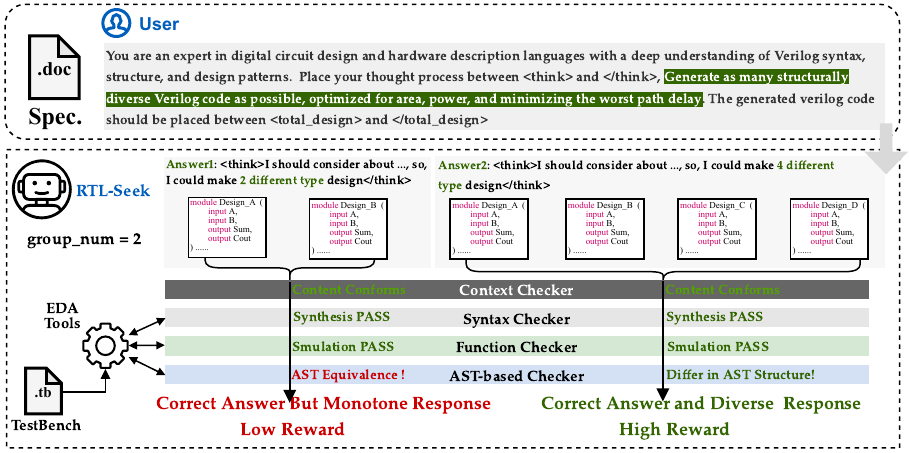}
    \vspace{-0.2cm}  
    \caption{Diversity-Centric Multi-Objective Reward reinforcement learning in RTLSeek.}
    \label{fig3}
    \vspace{-0.3cm}  
\end{figure*}

GRPO extends policy gradient methods by introducing group-wise relative comparisons, addressing critical limitations in conventional RL approaches for generative tasks. Originally demonstrated in DeepSeek R1\cite{deepseek-aiDeepSeekR1IncentivizingReasoning2025}, GRPO's unique characteristics make it particularly suitable for RTL generation:

\begin{itemize}[leftmargin=*]
    \item \textbf{Stabilized Group Updates}: By constraining policy updates within output groups through KL-divergence regularization, GRPO mitigates the mode collapse problem prevalent in PPO while maintaining exploration efficiency. The group-wise mechanism provides more reliable gradient estimates compared to single-sample methods.
    
    \item \textbf{Relative Quality Evaluation}: GRPO's novel optimization surface considers relative rankings within output groups, contrasting with PPO's absolute advantage estimation and DPO's pairwise comparisons. This enables joint evaluation of functionally equivalent but structurally diverse RTL designs.
\end{itemize}
Formally, given a question $q$, GRPO samples a group of outputs $\left \{ o_1,o_2,...,o_G \right \}$ from the old policy
$\pi_{\theta_{\mathrm{old}}}$ and optimizes the new policy $\pi_\theta$ by maximizing the objective function:

\begin{equation}
\begin{gathered}
\mathcal{J}_{\mathrm{GRPO}}(\theta) = \mathbb{E}\left[q\sim P(Q),\left\{o_i\right\}_{i=1}^G\sim\pi_{\theta_{\mathrm{old}}}(O[q])\right] \\
\frac{1}{G}\sum_{i=1}^G\left(\min\left(\frac{\pi_\theta(o_i|q)}{\pi_{\theta_{\mathrm{old}}}(o_i|q)}A_i,\mathcal{C}\right)-\beta\mathbb{D}_{KL}(\pi_\theta\|\pi_{ref})\right),\\
\mathcal{C} = \mathrm{clip}\left(\frac{\pi_\theta(o_i|q)}{\pi_{\theta_{\mathrm{old}}}(o_i|q)},1-\varepsilon,1+\varepsilon\right)A_i,
\end{gathered}
\end{equation}
where $\pi_{\theta}$ is the current policy, $\pi_{\theta_{\mathrm{old}}}$ is the old policy, $A_i$ is the advantage of the i-th output, and $\varepsilon$ is the hyperparameter used to control the update amplitude of the policy. $\beta $ is KL divergence penalty term coefficient, $\mathbb{D}_{KL} (\pi_\theta\|\pi_{ref})$ is KL divergence between the strategy $\pi_{\theta}$ and the reference strategies $\pi_{ref}$. 
\Cref{fig3} illustrates the group comparison mechanism with $G=2$, showing how relative evaluations enable diverse solution generation.
\subsubsection{Training Paradigm}

As shown in \Cref{fig2}, to maximize the utility of limited training data,
we develop a hybrid training paradigm combining SFT with a two-stage GRPO optimization process.
The complete training scheme enables simultaneous optimization for both correctness and diversity, overcoming the limitations of conventional approaches that typically prioritize one aspect at the expense of the other. The pipelines of the three stages are detailed as follows (Shown in \Cref{fig2}).



{Stage 1. SFT Warm-up: Learn what is HDL.}
We first train on curated RTL datasets containing both verified and unverified examples from reputable open-source projects. These natural language-to-Verilog pairs establish fundamental syntax and functional understanding, stabilizing later RL training. The SFT phase only requires a single Verilog module output for a query and focuses on learning basic code patterns and common design constructs, while filtering out low-quality examples through careful dataset curation. This initialization is crucial as it provides a strong starting point for subsequent reinforcement learning, preventing the model from exploring invalid design spaces during RL training.

{Stage 2. GRPO: Learn to generate diverse HDLs.}
Using large-scale unverified datasets, this phase employs diversity rewards to: (1) maximize data utility and generalization, while (2) developing varied yet syntactically valid RTL generation capability. We specifically design the diversity reward to encourage structural variations in control logic, datapath organization, and module hierarchy. The group-based sampling in GRPO allows the model to compare multiple design alternatives simultaneously, learning to generate different implementations for the same specification. This phase significantly expands the model's design repertoire beyond what could be learned through SFT alone.

{Stage 3. GRPO: Learn to generate right and diverse HDLs.}
The final stage applies combined diversity and functional rewards on verified datasets. This refines the model's functional-code correspondence understanding, requiring careful parameter calibration from previous phases. We implement a dynamic weighting scheme that automatically adjusts the balance between diversity and correctness rewards based on validation performance. The testbench verification provides precise feedback for functional correctness, while the diversity component maintains the variation learned in previous stages. This two-tiered reward structure ensures the model generates both correct and innovative designs.









\subsection{Multi-Objective Reward Scheduling}\label{subsec:met_mul}

{The key challenge} in improving the exploration quality of LLMs in RL lies in designing an appropriate reward mechanism that both refines decisions and provides systematic guidance throughout the learning process.

Traditional post-training paradigms based on SFT typically utilize syntax and functional correctness feedback from EDA tools~\cite{changDataAllYou2024a} to refine reference responses. However, these approaches often underutilize scarce training datasets and currently lack effective verification methods to ensure functional correctness for most testbench-free datasets.

As illustrated in Fig~\ref{fig3}, we propose Multi-Objective Reward Scheduling. This approach not only incorporates syntax and functional correctness feedback but also integrates a diversity reward obtained through Abstract Syntax Tree (AST) analysis to guide LLMs in generating structurally diverse RTLs. Additionally, we introduce a context reward to regulate reasoning quality, thereby achieving an effective exploration-exploitation balance.

The whole reward function \(R_{total}\) is as follows:

\begin{equation}
R_{total} = R_{syn} + R_{func} + R_{div} + R_{cont},
\end{equation}

Here, $R$ epresents the reward functions we have designed separately. By decomposing and combining different rewards for various RL training stages, we are able to maximize the exploration capabilities of the LLM within the constraints of limited training data.
These components are explained in detail below.


\subsubsection{Syntax Correctness Reward}

\(R_{syn}\) is obtained by verifying whether the generated Verilog code adheres to the syntactic rules of the Verilog language, using the syntax analysis tool, Pyverilog \cite{Takamaeda:2015:ARC:Pyverilog}: if any RTL in the generated set passes the syntax check, then \( R_{syn} = 1 \); otherwise, \( R_{syn} = 0 \).

\subsubsection{Function Correctness Reward}

\( R_{func} \) aims to guide the LLM in generating functionally correct RTL code. Only a small subset of datasets have a testbench verification set, which is used with the commercial simulation tool VCS to simulate the generated RTL. If any RTL in the generated set passes the testbench simulation, \( R_{func} = 1 \); otherwise, \( R_{func} = 0 \).

\subsubsection{Diversity Reward}

Since different RTL implementations can exist for the same functional specification under varying design objectives in IC design, we expect the LLM to explore as many diverse design solutions as possible, rather than repeatedly generating the same design. To achieve this goal, we propose 
\(R_{div}\) as follows:
\begin{equation}
R_{div}=
   N_{c}+ N_{s},
\end{equation}
where $N_c$ and $N_S$ represent the number of generated heterogeneous RTL codes that pass syntax check and functional verification, respectively.

\subsubsection{Context Reward}

\(R_{cont}\) consists of two components: format correctness and reasoning length reward. The format correctness requires the model to enclose the reasoning process of the given problem within the \(<think></think>\) tags, and to place the final multiple RTL solutions within the \(<total\_design></total\_design>\) tags. If the output match the format, a reward of 0.5 will be given; otherwise, a penalty of -0.5 will be given. The reasoning length reward is obtained from the significant correlation between the length of the reasoning response and the quality of the generated output~\cite{yeoDemystifyingLongChainofThought2025}. Specifically, if the reasoning response is too short, the model may fail to fully explain the reasoning steps, resulting in reduced accuracy and incompleteness in the generated response. Conversely, when the reasoning response is too long, the model may produce redundant or irrelevant content or become entangled in unnecessary, lengthy reasoning. To address this, we propose the context reward, which dynamically assigns rewards based on both the format of the answer and the length of the reasoning response. 
We define a satisfaction indicator \( I_{s}\) to reflect the quality of the current answer. If the sum of \(R_{syn}\), \(R_{func}\) and \(R_{div}\) is greater than a threshold \( \Delta = 4\), then \( I_{s} = 1 \); otherwise, \( I_{s} = 0 \).
Overall, the form of the reward function is as follows:

\begin{equation}
R_{cont}=\{\begin{matrix}
  0.5 L_t + 0.5I_f& \text{if } I_{s} = 1\\
 -0.5 L_t + 0.5I_f &\text{if } I_{s} = 0
\end{matrix}
\end{equation}

where, $I_f$ indicates whether the output meets the required format. If so, it is assigned a value of 1; otherwise, it is assigned -1, $L_t$ represents the ratio of the average length of the chains of thought from the previous four responses to the length of the current chain of thought.

\begin{figure}[tb!]
    \center
    \includegraphics[width=0.6\linewidth]{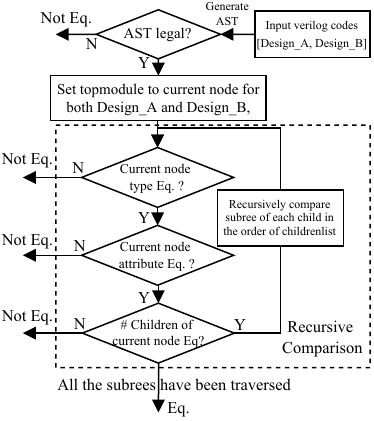}
    \vspace{-0.2cm}  
    \caption{Verilog structural equivalence testing algorithm based on abstract syntax tree.}
    \label{fig4}
        \vspace{-0.4cm}  
\end{figure}

\subsection{AST-Based Diversity Analysis}\label{subsec:met_ast}

To determine whether Verilog code segments differ substantively or merely in superficial aspects like variable naming or statement reordering, we implement equivalence verification based on AST, hierarchical representations that capture the syntactic structure of code while abstracting away lexical details such as whitespace and comments. For Verilog hardware description language, these tree structures effectively preserve module definitions, port declarations, signal assignments, and control structures in a form conducive to programmatic analysis \cite{takamaeda2015pyverilog}.

We parse source code into ASTs using pyverilog, then apply a recursive comparison algorithm that performs a depth-first traversal of both trees as shown in \Cref{fig4}.
This algorithm methodically compares node types, attributes, and structural relationships to determine complete equivalence. By operating at the structural level rather than the textual level, our approach overcomes limitations of text-based comparison by accurately identifying structural equivalence despite surface differences.

When code segments differ only in variable names (e.g., "p" versus "q"), our system recognizes their equivalence, while correctly distinguishing functionally different implementations even when they share certain syntactic patterns. This recursive verification system supports diversity exploration in RTL code generation for RTL-Seek by filtering out superficially different but structurally identical variants.

\section{Experiments}

\subsection{Experimental Settings}

\subsubsection{Dataset}

Due to limited high-quality Verilog data, we built our dataset by combining open-source sources from GitHub and Hugging Face. After preliminary cleaning with Design Compiler, 5167 synthesis-passed code-description pairs were selected for Stage 1 (SFT). For Stage 2, 3570 natural language descriptions were extracted from another subset. Stage 3 used 829 functionally verified samples from Verilog-eval \cite{liuVerilogEvalEvaluatingLarge2023} and similar sources.

\subsubsection{Training and Testing Setting}

We performed fine-tuning using {GRPOTrainer}, a reinforcement learning framework based on the \texttt{trl} library. 
To select a suitable base model, we ran pilot tests comparing different large language models on Verilog code generation. We ultimately chose {Qwen 2.5-7B-Instruct-1M} (abbreviated as Qwen 2.5) for its superior code-generation accuracy and output volume. All training was carried out on 8 NVIDIA A100 GPUs, each with 40\,GB of memory. We adopt the LoRA method for fine-tuning~\cite{huLoRALowRankAdaptation2021}, with the following hyperparameters: the rank of the LoRA layers is set to 4, the alpha parameter to 8, and the dropout rate to 0.1. During training, we use the Adam optimizer~\cite{kingma2017adammethodstochasticoptimization} with a learning rate of 5e-5.
We used all designs from the publicly available {RTLLM} benchmark~\cite{luRTLLMOpenSourceBenchmark2023} as our test set. The temperature for LLM inference is set to 0.2.


\subsubsection{Evaluation Metrics}

We denote syntax correctness as Syn., functional correctness as Fun., generation performance as Gen.,
and evaluate the model performance based on the following aspects:

\begin{itemize}
    \item \textbf{OPMO\_Pass@{\textit{k}}}: The traditional pass@{\textit{k}}, referred to in this paper as \textbf{One-Prompt-One-Output pass@{\textit{k}}} (OPOO\_Pass@{\textit{k}}), evaluates whether at least one of the Top-\textit{k} outputs is correct when each prompt generates a single Verilog module. To better assess the effectiveness of our approach, we introduce \textbf{One-Prompt-Multi-Output pass@{\textit{k}}} (OPMO\_Pass@{\textit{k}}), which evaluates whether any correct result exists among the multiple Verilog modules generated within the Top-\textit{k} outputs for a single prompt. Specifically, we only consider the cases of \textit{k} = 1 and \textit{k} = 5.

    \item {Success Rate:}    
    We use the proportion of correct RTL module code across all generated modules obtained from five queries as the Success Rate,~\Cref{result} and~\Cref{Abalation} show only functional Success Rate, detailed syntactical Success Rate could be found in Technical Appendix.
     \item {Quantitative Metrics:}  
    Gen.Num: The number of Verilog codes generated per prompt.
    Syn.Num: The number of syntactically valid codes per prompt.
    Fun.Num: The number of functionally correct codes per prompt.
\end{itemize}

\begin{figure}[tb!]
    \center
    \includegraphics[width=0.75\linewidth]{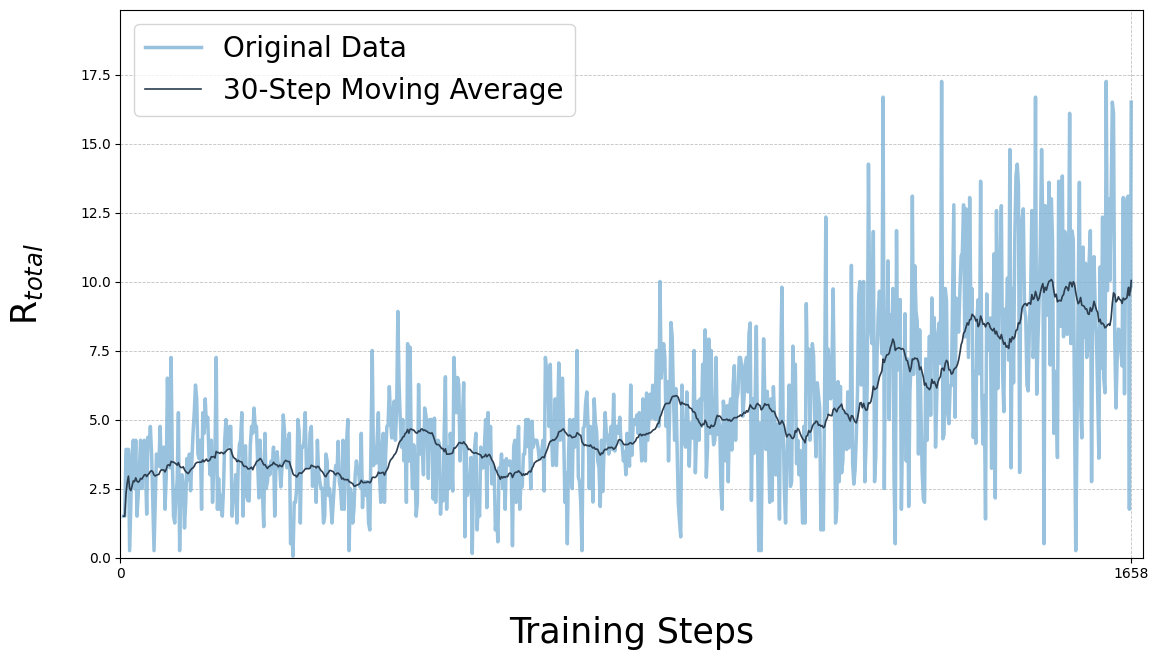}
        \vspace{-0.3cm}  
    \caption{The trend graph of the total reward obtained by every training step.}
        \vspace{-0.4cm}  
    \label{reward}
\end{figure}

\subsection{Evaluation Results}

\subsubsection{Training Dynamics of Reinforcement Learning}

\Cref{reward} shows the original data and the 30-step moving average of reward in Stage 3, where the overall trend exhibits a clear upward movement. This suggests that as training progresses, the model's rewards consistently increase, demonstrating the effectiveness of the training method.

\begin{table*}[tb!]
    \centering
    \caption{Generate result comparison about different metrics.}
    \vspace{-0.3cm}  
    \label{result}
    \makebox[\linewidth]{%
        \resizebox{.70\linewidth}{!}{
    \begin{tabular}{@{}l c c c c c c@{}}
        \toprule
        \textit{Type} & \textit{Model} &\textit{Params (\#)} &  \textit{Syn.{OPOO\_pass@1}} & \textit{Syn.{OPOO\_pass@5}} & \textit{Fun.{OPOO\_pass@1}} & \textit{Fun.{OPOO\_pass@5}}     \\ \hline
        \multirow{3}{*}{Foundation} & Qwen 2.5-Instruct*   &7B  & 0.48  &0.71  &0.27  &0.48   \\ 
        &GPT-4o*   &/  &0.80  & 0.89  & 0.42  & 0.66    \\ 
        &DeepSeek R1*   &671B  & 0.77  &0.86  &0.55  &0.73     \\
        \hline        
        \multirow{4}{*}{SFT models} & DeepRTL-2   &220M   & 0.71  &0.81  &0.32  &0.42  \\
        & DeepRTL-1   &16B  & 0.74  &0.77  &0.38 &0.35  \\
        & RTLCoder*   &7B & 0.73  &0.89  &0.32 &0.49  \\
        & Thakur*   & 16B & 0.83 & 0.86 & 0.17 & 0.24 \\ \hline
        \multirow{2}{*}{RL models}& ChipSeek-R1 & 7B & / &0.96 &/ & 0.83\\
        & CodeV-R1 & 7B & / & / &0.73 & 0.86 \\ \hline
        \textit{Type} & \textit{Model} & \textit{Params (\#)} &  \textit{Syn.{OPMO\_pass@1}} & \textit{Syn.{OPMO\_pass@5}} & \textit{Fun.{OPMO\_pass@1}} & \textit{Fun.{OPMO\_pass@5}}     \\
        \toprule
        \multirow{3}{*}{Foundation} & Qwen 2.5-Instruct*   &7B  & 0.50  &0.74  &0.32  &0.50   \\ 
        &GPT-4o*   &/  &0.86  & 0.93  &0.5  & 0.71    \\
        &DeepSeek R1*   & 671B  & \underline{0.90}  &0.96  &0.65 &0.83     \\ \hline  
        Ours & RTLSeek*   &7B  &  \textbf{0.86}  &\textbf{0.96}  &\textbf{0.76}  &\textbf{0.86} \\        
        \bottomrule
    \end{tabular}
   }}
    \begin{tablenotes}[flushleft]
    \footnotesize
    \item \textit{Note:} Models marked with an asterisk (*) were evaluated by ourselves; all others are cited from the respective papers; bold denotes the best result except commercial LLMs; underline denotes the real best result.
    \end{tablenotes}
 \vspace{-0.3cm}  
\end{table*}

\begin{table*}[tb!]
    \centering
    \caption{Ablation study on the diversity reward and multi-stage training.}
    \vspace{-0.3cm}  
    \label{Abalation}
        \resizebox{.65\linewidth}{!}{
    \begin{tabular}{@{}lp{1cm}p{1cm}p{1cm}p{1.6cm}p{1.6cm}p{1.6cm}p{1.6cm}p{1.6cm}@{}}
        \toprule
        Ablation  & Gen. Num & Syn. Num & Fun. Num & Syn.OPMO \_pass@1  & 
        Syn.OPMO \_pass@5  & Fun.OPMO \_pass@1 & Fun.OPMO \_pass@5  & Success Rate \\
        \toprule
        RTLSeek    &3.22  &\textbf{2.23}  &\textbf{1.27}  &\textbf{0.86}  & \textbf{0.96} &\textbf{0.76}  &\textbf{0.86} &\textbf{0.40}  \\
        RTLSeek.w/o DR    &1.70  &0.85  &0.54  &0.64 &  0.73 &0.55  &0.59 &0.35  \\
        RTLSeek.w/o S3    &\textbf{4.14}  &1.31  &0.61  &0.41 &  0.73 &0.32  &0.55 &0.15   \\
        RTLSeek.w/o S2    &2.42  &1.33  &0.67  &0.55 & 0.77  &0.45  &0.59  &0.28 \\
        RTLSeek.only S1 &  1.15 & 0.32 & 0.17 & 0.32 & 0.59 & 0.23 & 0.31 & 0.15\\
        \bottomrule
    \end{tabular}
    }
    \vspace{-0.3cm}  
\end{table*}

\subsubsection{Overall Evaluation on RTLLM}


We evaluate RTLSeek on RTLLM v1.1 in~\Cref{result} against both commercial foundation models--Qwen2.5-Instruct~\citep{yang2025qwen2}, GPT-4o~\citep{openaiGPT4TechnicalReport2024}, and DeepSeek-R1~\citep{deepseek-aiDeepSeekR1IncentivizingReasoning2025}--and open-source baselines, including the supervised-fine-tuned (SFT) DeepRTL~\citep{liuDEEPRTLBRIDGINGVERILOG2025}, Thakur~\citep{thakurVeriGenLargeLanguage2024}, and RTLCoder~\citep{liu2024rtlcoderfullyopensourceefficient}, as well as RL post-training academic works like ChipSeek-R1~\citep{chen2025chipseekr1generatinghumansurpassingrtl} and CodeV-R1~\citep{zhu2025codevr1reasoningenhancedveriloggeneration}.
For foundation models we apply two prompt-engineering strategies: OPOO and OPMO.
For the academic fine-tuned models (Thakur and RTLCoder) we report results with OPOO prompting and the corresponding OPOO\_pass@k metric; OPMO prompting fails on these models because domain-specific fine-tuning degrades instruction-following ability—a manifestation of catastrophic forgetting that prevents generation of multiple correctly-formatted RTL modules.
All other entries in the table are taken directly from the respective papers; however, there is every reason to expect that the same catastrophic-forgetting effect would occur in those models as well. Restricting the comparison to OPOO prompting is therefore fully justified.
As shown in \Cref{result}, RTLSeek raises the average functional success rate by 29\% over Qwen2.5, despite using only 1\% of R1’s parameters.
Its enhanced diversity yields more correct outputs per prompt, substantially improving OPMO\_pass@5 and overall functional success rate.




 

\subsection{Ablation study}

\subsubsection{Study Settings and Model Descriptions}

To verify our hypotheses regarding which methods can effectively enhance the Verilog generation capabilities of LLMs, we conducted an ablation study. The ablated models we implemented include:
(1) {RTLSeek.w/o DR}: RTLSeek without the Diversity Reward in Stage 2 and 3 RL training.
(2) {RTLSeek.w/o S3}: RTLSeek without Stage 3 RL training, leaving only Stages 1 and 2.
(3) {RTLSeek.w/o S2}: RTLSeek without Stage 2 RL training; after Stage 1, the process directly moved to Stage 3.
(4) {RTLSeek.only S1}: RTLSeek only after Stage 1 OPOO SFT training.

\begin{table}[h]
    \centering
        \vspace{-0.2cm}  
    \caption{Fine-grained ablation study on two subsets obtained from the partitioning RTLLM v1.1 for Success Rate about the diversity reward.}
    \label{correct}
    \label{Abalation_2}
            \resizebox{.75\linewidth}{!}{
    \begin{tabular}{@{}lp{2.5cm}p{3cm}@{}}
        \toprule
        Model   & w/o DR Pass & w/o DR NO Pass  \\
        \toprule
        RTLSeek     &0.56  &0.131    \\
        RTLSeek.w/o DR     &0.47  &0   \\

        \bottomrule
    \end{tabular}
    }
        \vspace{-0.3cm}  
\end{table}
\subsubsection{Ablation Study for Diversity Reward}

As indicated in \Cref{Abalation}, RTLSeek outperforms RTLSeek. w/o DR in terms of {Success Rate}.
This comparison suggests that diversity-oriented RL training is important for improving LLM-based RTL generation.
In addition, RTLSeek also shows a significant increase in average Fun.{OPMO\_pass@5} when compared to RTLSeek. w/o DR. This improvement mainly stems from the diversity-oriented approach, which encourages the model to generate more varied responses per prompt, expanding the solution space and enhancing overall performance.
We also conducted a more detailed study in~\Cref{Abalation_2}.
In~\Cref{Abalation_2}, we divide the items of RTLLM v1.1 benchmark into two categories based on whether the model {RTLSeek.w/o DR} can pass:
(1) w/o DR Pass: Items that {RTLSeek.w/o DR} can solve correctly at least once within five attempts.
(2) w/o DR NO Pass: Items that {RTLSeek.w/o DR} can not solve (no correct solutions even after five attempts).
In \Cref{Abalation_2}, we report the average Success Rate of RTLSeek and RTLSeek.w/o DR on these two sets of items.
(1) For items w/o DR NO pass, RTLSeek.w/o DR’s average Success Rate is 0\%, but Ours achieves 13.1\%.
(2) For items without DR pass, RTLSeek.w/o DR has 47\% average Success Rate, while RTLSeek achieves 56\%, surpassing Baseline by 19\%.
These results demonstrate that introducing a diversity reward not only boosts correctness on items that RTLSeek.w/o DR can already solve but also enables solving some RTLSeek.w/o DR unsolved items.
{Hence, diversity-oriented reinforcement learning training has a notable and measurable impact on enhancing LLM-based RTL code generation.}

\subsubsection{Ablation Study for Multi-Stage RL Training}


As shown in \Cref{Abalation}, RTLSeek improves average Success Rate by 166.7\% over \textit{RTLSeek w/o S3}, confirming the strong benefit of Stage~3, which uses testbench-equipped data for diversity-driven exploration and correctness learning. RTLSeek also exceeds \textit{RTLSeek w/o S2} by 42.9\%, with notably more generated codes, showing that Stage~2 enhances diversity and enables more effective Stage~3 exploration. Overall, the Diversity-Centric Multi-Objective Reward Scheduling and three-stage framework allow RTLSeek to surpass GPT-4o in both RTL syntax and functional accuracy, with ablations validating the effectiveness of the diversity-oriented RL paradigm.

\section{Conclusion}


This paper presents RTLSeek, a novel post-training paradigm using Diversity-Oriented Reinforcement Learning to improve the accuracy and diversity of LLM-generated RTL. With a Diversity-Centric Multi-Objective Reward and a three-stage training framework, RTLSeek effectively addresses data scarcity. Experiments show it outperforms GPT-4 and DeepSeek R1 on RTLLM benchmark, with ablation studies confirming its effectiveness.

\newpage

\bibliographystyle{ACM-Reference-Format}
\bibliography{ref}










\end{document}